\documentclass[aps,prb,twocolumn,
superscriptaddress,
groupedaddress,10pt]{revtex4}
\usepackage{amsmath}
\usepackage{amssymb}
\usepackage{graphicx}
\usepackage{epsfig}
\usepackage{dcolumn}
\usepackage{bm}
\usepackage{bm}
\usepackage{blindtext}
\usepackage{natbib}

\usepackage{bbm}
\usepackage{multirow}

\usepackage{mathtools}

\def\be{\begin{equation}} \def\ee{\end{equation}}
\def\bea{\begin{eqnarray}} \def\eea{\end{eqnarray}}

\def\nn{\nonumber}

\begin{document}
\title{
Superconducting junction with  tri-component pairing gap functions
}
\author{Chao Xu}
\affiliation{Department of Physics, University of California,
San Diego, California 92093, USA}
\author{Wang Yang}
\affiliation{Department of Physics and Astronomy and Stewart Blusson Quantum Matter Institute,
University of British Columbia, Vancouver, B.C., Canada, V6T 1Z1}
\author{Congjun Wu}
\affiliation{Department of Physics, University of California,
San Diego, California 92093, USA}

\begin{abstract}
We study a superconducting hetro-junction with one
side characterized by the unconventional chiral $p$-wave gap function $p_x\pm ip_y$
and the other side the conventional $s$-wave one.
Though a relative phase of $\pm \frac{\pi}{2}$ between any two components
of gap functions is favored in the junction region,
mutual phase differences cannot achieve $\pm \frac{\pi}{2}$ simultaneously,
which results in frustration.
Based on a Ginzburg-Landau free energy analysis,
the frustrated pattern is determined to be $s+ i\eta_1 (e^{ i\eta_2 \varphi/2}p_x +\eta_3 e^{- i\eta_2 \varphi/2}p_y)$ with
$\eta_j=\pm 1$ ($j=1,2,3$), where $\varphi$ is the phase difference
between the $p_x$- and $p_y$-wave gap functions.
Furthermore, we  find that
the junction exhibits an anisotropic magnetoelectric effect,
manifesting itself as an anisotropic spin magnetization
along the edge of the junction.
\end{abstract}

\maketitle

\section{Introduction}

Chiral superconductors constitute a class of superconducting states of matter
characterized by unconventional gap functions, spontaneous time-reversal symmetry
breaking, and nontrivial topological properties \cite{Kallin2016}.
The topological structure in the pairing wavefunctions
leads to exotic phenomena, including  the  emergence of Majorana zero modes in vortex cores \cite{Read2000,DasSarma2006,Fu2008,Sau2010,Teo2010} and
chiral Majorana fermions on the boundary of the system \cite{Yu2010,Qi2010,Chung2011},
which can be useful in realizing topological quantum computations \cite{Kitaev2003,Kitaev2006,Stone2006,Alicea2011,Halperin2012}.
The superconducting
$\rm{Sr_2\rm{Ru}O_4}$ \cite{Maeno1994,Mackenzie2003,Maeno2012,Liu2015} and $\rm{UPt_3}$ materials \cite{Joynt2002,Schemm2004,Strand2009,Avers2020} have been proposed to host chiral superconductivity with $p$- and $f$-wave pairing gap functions, respectively,
though there are still debates over the pairing nature of these materials \cite{Maeno2012,Kallin2009,Kallin2012,Mackenzie2017}
despite intensive theoretical and experimental studies \cite{Ishida1998,Duffy2000,Laube2000,Mackenzie1998,Luke1998,Nelson2004,Xia2006,Kidwingira2006,Pustogow2019}.

In general, when instabilities in several pairing channels coexist,
the system may develop a superposition of gap function symmetries which
spontaneously breaks time-reversal symmetry.
A typical pattern of time-reversal symmetry breaking is that a relative
$\pm\frac{\pi}{2}$ phase difference develops between two different
pairing channels with different symmetries, which has been studied in
various systems including the $\rm{^3He}$-A superfluid phase
\cite{Volovik1988, Volovik1989}, and superconductors with $p_x$+$ip_y$ \cite{Kopnin1991,Ivanov2001,Stone2006,Tewari2007,Chuanwei2008,Fu2008,
Cheng2010,Qi2009}, and $d_{x^2-y^2}$+$id_{xy}$\cite{Laughlin1998,Senthil1999,Horovitz2003,Hu2008,Sato2010,
Black2012,Chubukov2012,Wang2012,Kiesel2013,Liu2013,Black2014,Liu2018,Kennes2018,Yang2018,Huang2019}
gap function symmetries.
The mixing between the $s$-wave and $p$-wave gap function symmetries with
a relative phase difference $\pm\frac{\pi}{2}$ was first proposed
by one of the author and Hirsch in the context of superfluid
instability of dipolar fermions \cite{Wu2010}, and was later
generalized to other systems \cite{Wang2014,Wang2017,Yang2017}.
Mixed gap function symmetries breaking time-reversal symmetry have also
been proposed in the iron-based superconductors \cite{Lee2009,Hu2020}
and other related systems,
such as $s$+$id$\cite{Lee2009,Thomale2011,Platt2012,Khodas2012,Fernades2013,
Hinojosa2014,Lin2016},
$s$+$is$\cite{Stanev2010,Lin2012,Marciani2013,Maiti2013,Ahn2014,
Garaud2014,Maiti2015,Lin2016,Hu2020}.
On the other hand, the interplays among three or more different
superconducting order parameters remain less explored
\cite{Garaud2011,Garaud2013,Lin2014,Yerin2014,Yerin2017}.

\begin{figure}
\includegraphics[width=6cm]{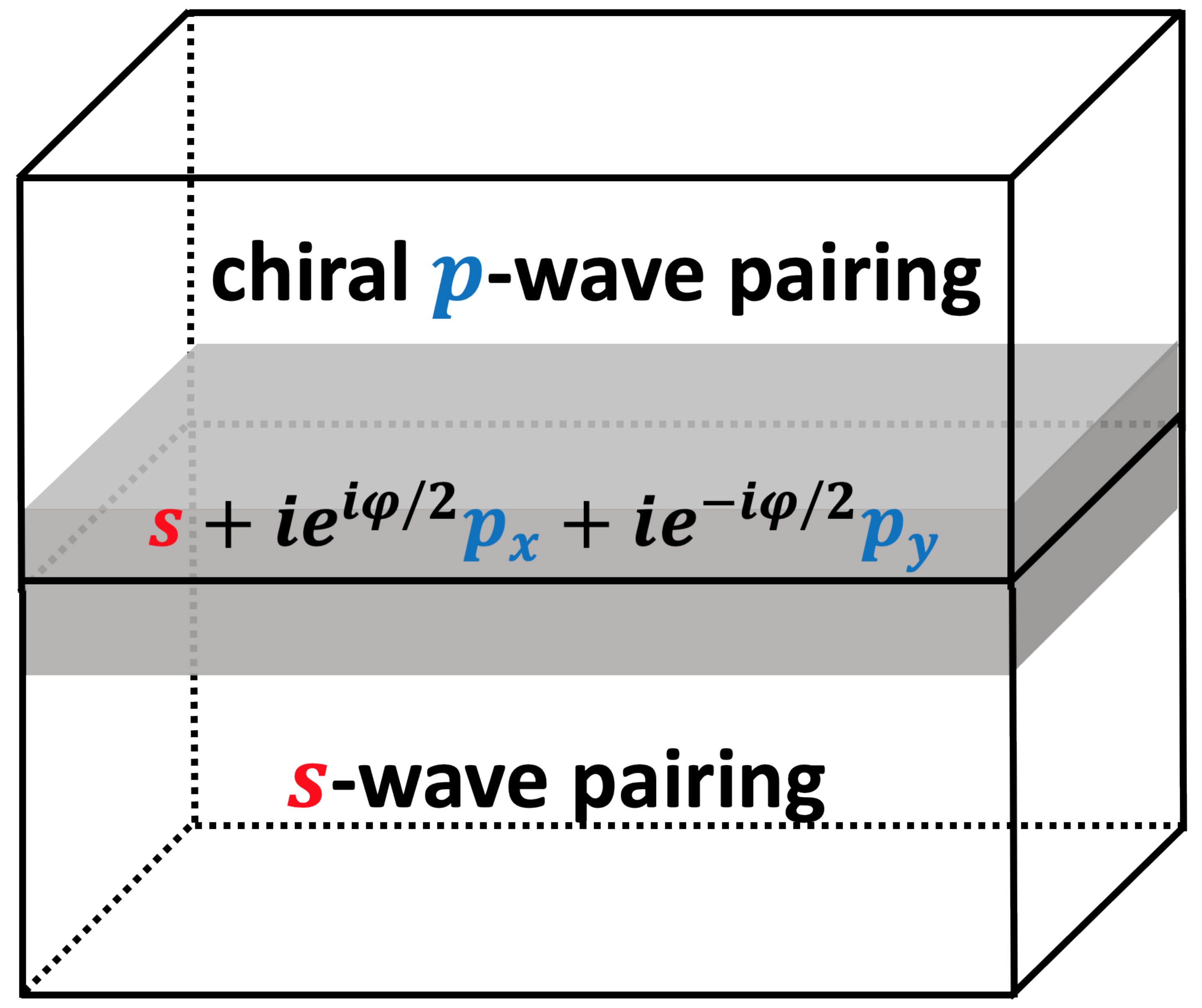}
\caption{
The heterojunction formed by a chiral $p$-wave superconductor in the upper 
space and an $s$-wave superconductor in the lower space.
A mixed tri-component gap function develops near the interface of the 
heterojunction induced by the proximity effect.
The $z$-direction is chosen along the crystalline $c$-axis as pointing upwards.
}
\label{set}
\end{figure}

\begin{figure}
\includegraphics[width=8.6cm]{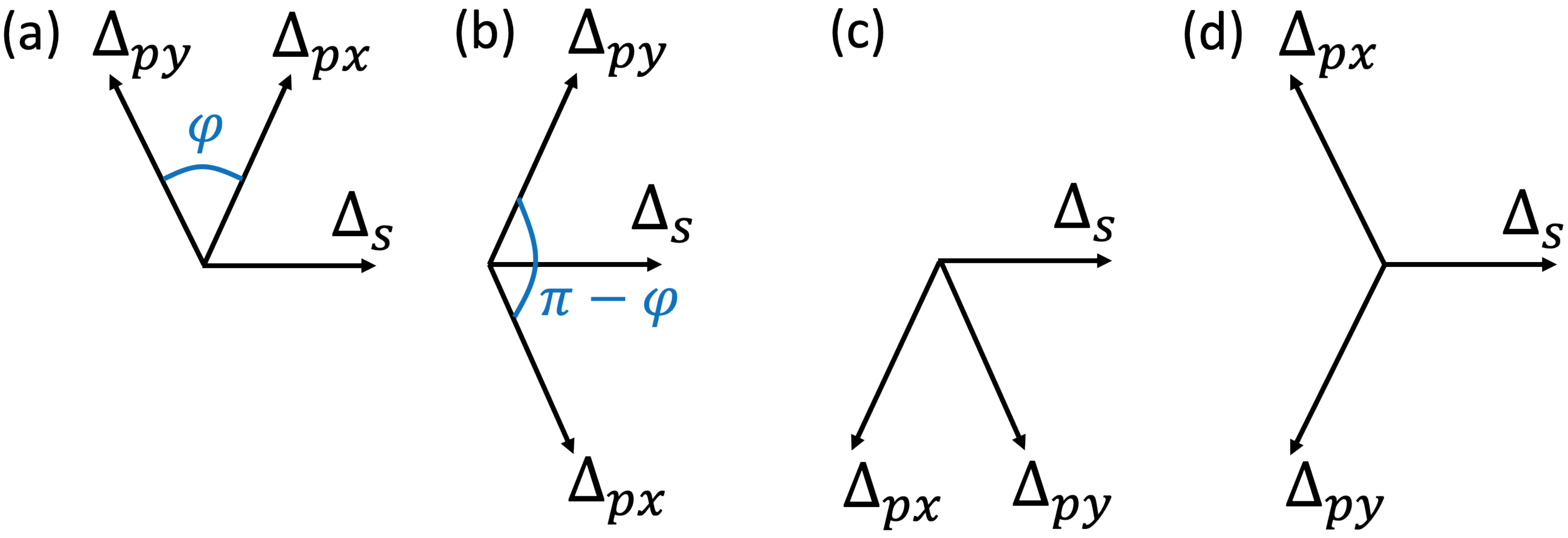}
\caption{ Plots of the four tri-component pairing configurations with the positive chirality, i.e.,  $s+i\eta_1 (e^{i\eta_2 \varphi/2}p_x+\eta_3 e^{-i\eta_2 \varphi/2}p_y)$ where $\eta_2\eta_3=-1$.
In ($a$-$d$), $|\Delta_{px}|=|\Delta_{py}|$,
and the phase of the $s$-wave pairing is fixed to be zero.
The configurations in ($b$), ($c$), and ($d$) can be obtained by performing
the $C_4$ rotations at the angles of  
$\pi/2$, $\pi$, and $3\pi/2$  on the configuration of ($a$), respectively.
We note that the rotations are performed in the orbital space not on the phase
configurations of the gap functions illustrated in Figures ($a$-$d$).
Hence, 
the rotation of $\pi/2$ keeps $\Delta_s$ unchanged, and 
$\Delta_{px} \to \Delta_{py}$ and $\Delta_{py}\to -\Delta_{px}$.
}
\label{tri}
\end{figure}

In this article, we study the superconductor-superconductor
junction with one side characterized by a chiral $p$-wave gap
function symmetry and the other side  the conventional $s$-wave
one, respectively, as illustrated in Fig. \ref{set}.
In the junction region, three gap function symmetries coexist
due to the proximity effect. 
The linear Josephson coupling is not allowed due to their 
different symmetries, and any two of them can only be coupled
via the quadratic Josephson term at the lowest order. 
Any two of them favor a relative phase of $\pm \frac{\pi}{2}$,
however, the system is frustrated since a simultaneous mutual
$\pm\frac{\pi}{2}$ phase difference is impossible among three
order parameters.
This frustration is different from that of the antiferromagnetism 
defined in the triangular lattice in which the bilinear Heisenberg 
coupling is analogous to the linear Josephson coupling. 
To determine the frustrated pattern of the gap functions, a Ginzburg-Landau
free-energy analysis is performed.
The gap function structure in the junction region is solved to
exhibit an exotic form
$s+ i\eta_1 (e^{ i\eta_2 \varphi/2}p_x+\eta_3 e^{- i\eta_2 \varphi/2}p_y)$
as shown in Fig. \ref{tri}, where $\varphi$ is the phase difference
between the $p_x$- and $p_y$-pairing order parameters,
and $\eta_j=\pm 1$ ($j=1,2,3$).
By fixing the chirality deep in the $p$-wave layer as the boundary condition,
the time-reversal and reflection symmetries are explicitly broken.
The frustration  spontaneously breaks the $C_4$ symmetry and can be viewed as
a frustration induced nematic superconductivity.
In the junction region, the tri-component pairing further breaks the
residual $C_4$ symmetry, and the four degenerate configurations satisfy $\eta_2\eta_3=-\eta_c$ ($\eta_c=\pm 1$) when the boundary condition is
chosen as $p_x+i\eta_c p_y$.

Furthermore, we find that the system exhibits an anisotropic
magnetoelectric effect around the edge of the junction, consistent
with the $C_4$ symmetry breaking.
The magnetoelectric effect also manifests itself as the emergence of an
anisotropic spin magnetization on the edge of the junction, which can
be analyzed through the splitting of the two spin-polarized chiral
Majorana edge modes.

The rest part of this article is organized as follows.
In Sect. II, the Ginzburg-Landau free energy analysis is performed,
and the origin of frustration among gap functions is illustrated.
The anisotropic magneto-electric effect and the edge magnetization
are studied in Sect. III.
The relation between the edge magnetization and the gap function
mixing is presented in Sect. IV.
Conclusions are given in Sect. V.

\section{Ginzburg-Landau free energy analysis}

\subsection{Brief review of the $p_x\pm ip_y$ pairing}

We first briefly review the Ginzburg-Landau free energy analysis for the chiral $p$-wave superconductor with the $p_x\pm ip_y$ pairing.
The point group symmetry is assumed to be the $D_{4h}$ group,
which applies to a tetrahedral lattice system.
The most general Ginzburg-Landau free energy respecting the U(1) gauge, the time reversal,  and the $D_{4h}$ point group symmetries up to quartic order is
\begin{eqnarray}
f_1 &=& \alpha_p(|\Delta_{px}|^2 +|\Delta_{py}|^2)- g_{pp} |\Delta^{*}_{px}\Delta_{py} -\Delta^{*}_{py}\Delta_{px}|^2\nonumber\\
&&+ \beta_p (|\Delta_{px}|^2 + |\Delta_{py}|^2)^2 + \beta^{\prime}_{p}(|\Delta_{px}|^4 + |\Delta_{py}|^4),
\label{eq:f_1}
\end{eqnarray}
in which $\Delta_{px}$, $\Delta_{py}$ are the order parameters of the $p_x$- and $p_y$-wave pairing gap functions, respectively;
$\alpha_p<0$ in the superconducting state;
$\beta_p>0$ is the coefficient of the corresponding rotationally invariant phase-non-sensitive quartic term;
the $\beta^{\prime}_{p}$  term breaks the SO(2) rotational symmetry  down to $C_4$;
$g_{pp}>0$ is the coefficient of the term which contains the quadratic Josephson coupling $(\Delta_{px}^*\Delta_{py})^2+\text{h.c.}$; and only the uniform parts of the free energy are kept while the gradient terms are neglected.

Since $g_{pp}$ is generically positive,
the energy of the quadratic Josephson  term is lowered if a $\pm\pi/2$ phase
difference is developed between $\Delta_{px}$ and $\Delta_{py}$.
As a result, the $p_x\pm ip_y$ pairing is favored which spontaneously
breaks the time-reversal symmetry.
Though the $p_x\pm ip_y$ pairing breaks both U(1) gauge and $C_4$ rotational symmetries,
it is invariant under $G R(\hat{z},\pi/2)$,
where $R(\hat{z},\pi/2)$ is the $\pi/2$ rotation around z-axis in the orbital space and $G$ is the gauge transformation by $\pm\pi/4$ of the electrons (i.e., $\pm\pi/2$ phase rotation of the Cooper pairs).
In particular, $L_z+\frac{1}{2}N$ remains to be a conserved quantity when $\beta_p^\prime=0$.

\subsection{Minimization of the free energy for the junction}

\begin{figure*}
\includegraphics[width=14cm]{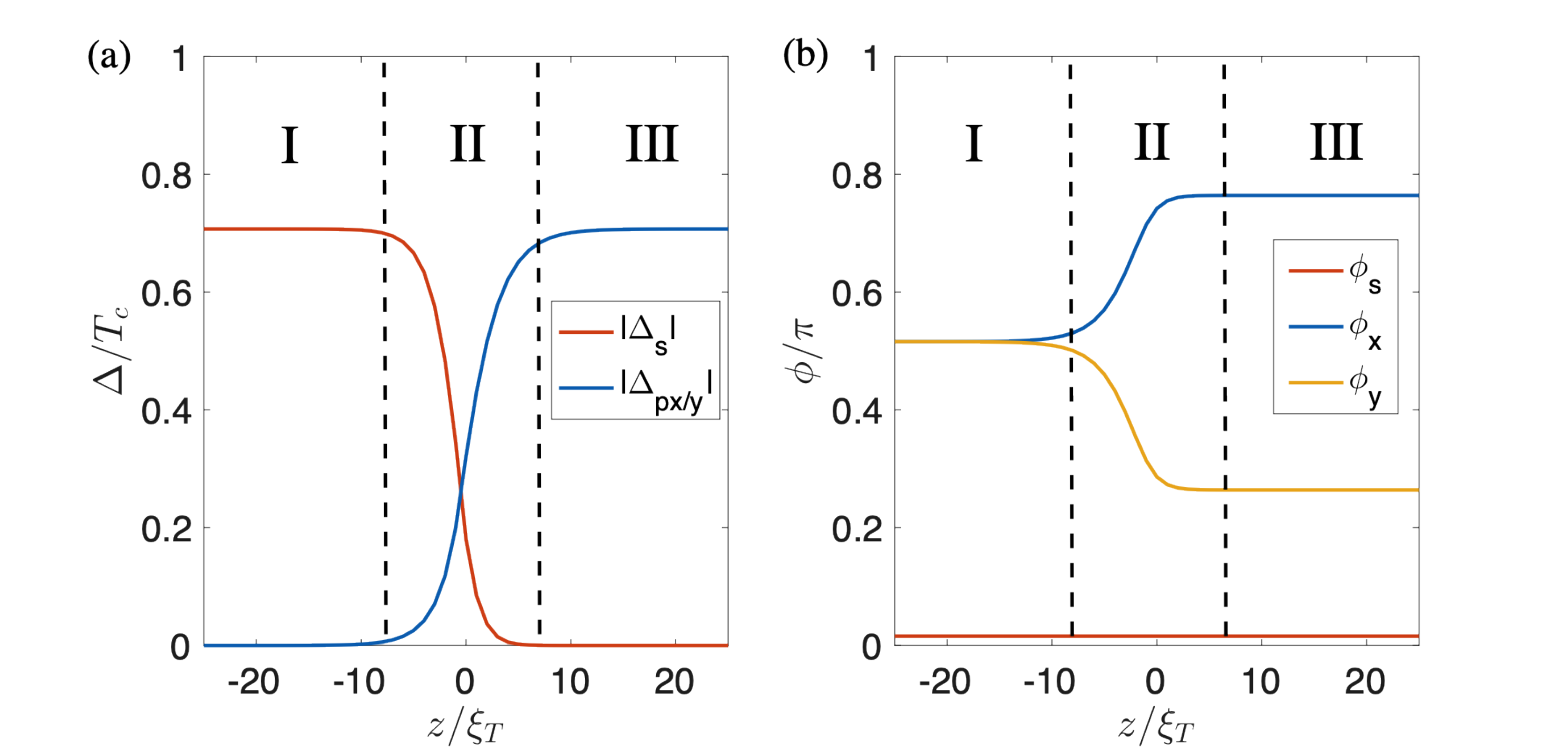}
\caption{
Magnitudes of the gap function $|\Delta_s|$ (red curve) and $|\Delta_{px}|=|\Delta_{py}|$ (blue curve) as functions of $z$ in ($a$);
and their phases $\phi_s$ (red curve), $\phi_x$ (blue curve) and $\phi_y$ (yellow curve) as functions of $z$ in ($b$).
The intervals of $z$ marked by ``I, II, III" represent the regions where 
$s$-wave dominates, $s$- and $p$-wave coexist, and $p$-wave dominates, 
respectively.
The units for $|\Delta_\lambda|$ ($\lambda=s,p_x,p_y$) and the spatial 
coordinate $z$ are the transition temperature $T_c$ and the coherence
length $\xi_T=v_F/T_c$, respectively, where $v_F$ is the Fermi velocity.
The parameters in the numerical calculations are chosen as 
$K_s=K_p=10N_F\xi_T^2$,
$\alpha_s = 2 \text{sgn}(z)N_F$,  $\alpha_p = -\text{sgn}(z)N_F$,
$\beta_s = 2N_F/T_c^2$,
$\beta_p = 3.75N_F/T_c^2$,
$\beta_p^{\prime} = 0.5N_F/T_c^2$,
$g_{sp} = 3.5N_F/T_c^2$,
$g_{pp} = 3.5N_F/T_c^2$,
$\gamma = 10N_F/T_c^2$,
where $N_F$ is the density of states at the Fermi level,
and $T_c$ is the superconducting transition temperature.
}
\label{mag}
\end{figure*}

Next we proceed to discuss the tri-component pairing gap function as a consequence of the competition among three pairing order parameters.
The system under consideration is a heterojunction formed by a chiral $p$-wave superconductor in the upper 
space and an $s$-wave superconductor in the lower space,
as shown in Fig.~\ref{set}.
The pairing Hamiltonians deep in the upper and lower spaces in Fig.~\ref{set} are given by
\bea
\hat{\Delta}_p&=&\sum_{\vec{k}\alpha\beta}  \frac{1}{k_f}(|\Delta_{px}|k_x +i |\Delta_{py}|
k_y)(\sigma^z i\sigma^y)_{\alpha\beta}c_{\vec{k}\alpha}^\dagger c^\dagger_{-\vec{k}\beta},\nn\\
\hat{\Delta}_s&=&|\Delta_s|\sum_{\vec{k}}c^\dagger_{\vec{k}\uparrow}c^\dagger_{-\vec{k}\downarrow},
\label{eq:Delta_pairing}
\eea
in which $\alpha,\beta=\uparrow,\downarrow$ are the spin indices; $c^\dagger_{\vec{k}\alpha}$ is the electron creation operator with momentum $\vec{k}$ and spin $\alpha$.

On the other hand,
due to the proximity effect, there is a mixture of $p$-wave ($\Delta_{px},\Delta_{py}$) and
$s$-wave ($\Delta_s$) superconducting order parameters in the junction region.
To study the pattern of the mixture,
 we take a Ginzburg-Landau free energy analysis.
Because of the heterostructure,
the point group symmetry becomes the planar $C_{4v}$ group, which contains the $C_4$ rotations and four reflections.
Assuming $U(1)$ gauge, time reversal, and $C_{4v}$ symmetries,  the free energy
density up to quartic order takes the form
\bea
f = f_{p} + f_{s} + f_{sp} + f_{sp}^{\prime},
\label{eq:f_pxpys}
\eea
in which
\begin{eqnarray}
	f_{s} &=& K_s |\nabla_z \Delta_s|^2 + \alpha_s |\Delta_s|^2 + \beta_s |\Delta_s|^4 \nn\\
	f_{p} &=& K_p( |\nabla_z \Delta_{px}|^2+|\nabla_z \Delta_{py}|^2\big) +f_1\nn\\
	f_{sp} &=& g_{sp} [\Delta^{*2}_{s} (\Delta^2_{px}+\Delta^2_{py})+\text{c.c.}]\label{fsp}\nn\\
	f_{sp}^{\prime} &=& \gamma (|\Delta_{px}|^2 +|\Delta_{py}|^2) |\Delta_s|^2,
	\label{eq:fs}
\end{eqnarray}
where $f_1$ within $f_p$ is given by Eq. (\ref{eq:f_1})
and ``c.c." is ``complex conjugates" for short.
The coefficient of each term up to tree level can be determined by a diagrammatic calculation as discussed in detail in Appendix \ref{app:GL}.

To mimic the junction structure close to the $z=0$ interface,
we set
\bea
&\alpha_p(z)<0, \alpha_s(z)>0,~ \text{for} ~z>0,\nn\\
&\alpha_p(z)>0, \alpha_s(z)<0,~ \text{for} ~z<0,
\eea
so that the $p_x+ip_y$ pairing dominates deep in the upper space,
whereas the $s$-wave pairing dominates deep in the lower space.
Due to the gradient terms led by $K_p$ and $K_s$, the pairing gap function
cannot exhibit a sudden change.
Therefore we expect that the $p_x$-, $p_y$- and $s$-wave pairing symmetries should coexist close to the  $z=0$ interface.

To obtain an intuitive understanding, we take a quick look at the phase sensitive terms in the free energy.
The phase sensitive $g_{sp}$ and $g_{pp}$ terms are
\bea
g_{sp} [\Delta^{*2}_{s} (\Delta^2_{px}+\Delta^2_{py})+\text{c.c.}]
 - g_{pp} |\Delta^{*}_{px}\Delta_{py} - \Delta^{*}_{py}\Delta_{px}|^2,
\eea
which can be evaluated as
\begin{flalign}
&2g_{sp}|\Delta_s|^2|\Delta_p|^2[\cos(2\phi_x-2\phi_s)+\cos(2\phi_y-2\phi_s)]\nn\\
&~~~~~~~+2g_{pp}[\cos(2\phi_x-2\phi_y)-1],
\label{eq:f_cos}
\end{flalign}
 where $\Delta_s=|\Delta_s|e^{i\phi_s}$, $\Delta_{px} = |\Delta_p|e^{i\phi_x}$, and $\Delta_{py} = |\Delta_p|e^{i\phi_y}$.
Each term in  Eq. (\ref{eq:f_cos}) is minimized if $\phi_{x}$, $\phi_{y}$ and $\phi_s$ mutually differ by $\pm\pi/2$.
However, Eq. (\ref{eq:f_cos}) is frustrated since a simultaneous mutual $\pi/2$ difference among three phases is impossible.
Therefore, there will be a competition between the phases of the superconducting order parameters in the coexisting region.

To determine the pattern arising from the competition, we apply an iterative numerical method to obtain the solution of the pairing gap function by minimizing the free energy.
The numerical results for the magnitudes and phases of the superconducting order parameters are displayed in Fig. \ref{mag} (a) and (b), respectively.
It is found that the solutions of the magnitudes $|\Delta_{px}|$ and $|\Delta_{py}|$ are equal as shown in Fig. \ref{mag} (a).
As can be seen from Fig. \ref{mag} (a), the system can be clearly divided into three regions:
the region  marked with ``I" where the $s$-wave pairing dominates (deep inside the $s$-wave bulk);
region ``II" between the two vertical dashed lines
where all the three pairing symmetries coexist;
and region ``III" where the $p_x,p_y$-wave pairings dominate (deep inside the bulk of the chiral $p$-wave superconductor).
In the numerical calculations, the phase $\phi_s$ of the $s$-wave pairing is chosen to be zero for $z<0$ and $|z/\xi_w|\gg1$ where $\xi_w=\sqrt{|K_s/\alpha_s|}$ represents the width of the coexisting region.
Then $\phi_s$ is solved to remain at zero in the entire junction as indicated by the red line in Fig. \ref{mag} (b).

As can be seen from Fig.~\ref{mag} (b), deep inside the $p$-wave bulk,
$\Delta_{px}$ and $\Delta_{py}$ have a relative $\pi/2$ phase difference,
and the magnitude  of $\Delta_s$ is nearly negligible.
When approaching the junction from the $p$-wave side,
the magnitudes of $\Delta_{px}$ and $\Delta_{py}$ start
shrinking and so does the phase $\varphi$ between them, whereas the magnitude of $\Delta_s$ keeps growing.
Eventually when leaving the coexisting region and entering the $s$-wave bulk, $\Delta_s$ is much larger than
$\Delta_{px}$ and $\Delta_{py}$ in magnitude.
We note that the three phases $\phi_{x}$, $\phi_{y}$ and $\phi_s$ exhibit the following pattern throughout the whole space,
\bea
\phi_x-\phi_y &=& \varphi\nn\\
\frac{\phi_x+\phi_y}{2}-\phi_s&=&\frac{\pi}{2}.
\eea
As a result, the tri-component pairing gap function in the coexisting region can be written as $s+i(p_xe^{i\varphi/2}+p_ye^{-i\varphi/2})$ as shown in Fig. \ref{tri} (a),
in which $\varphi$ decreases from $\pi/2$ down to $0$ as the junction is traversed from $z>0$ to $z<0$.

\subsection{Symmetry breaking pattern}

In closing this section, we discuss the symmetry breaking pattern in the junction region.
Clearly, all symmetry transformations $T,C_4,M_x,M_y,M_{x-y},M_{x+y}$ are spontaneously broken, where $M_{f(x,y)}$ represents the spin-orbit coupled reflection with respect to the $f(x,y)=0$ plane.
In particular, $L_z+\frac{1}{2}N$ is not conserved when $\beta_p^\prime=0$.
However, the tri-component pairing $s+i(p_xe^{i\varphi/2}+p_ye^{-i\varphi/2})$ is invariant under  $TM_{x-y}$.
Hence, the unbroken symmetry group is $\langle TM_{x-y}\rangle\simeq \mathbb{Z}_2$,
in which $\langle\cdot\cdot\cdot\rangle$ represents a group generated by the operations inside the bracket.
As a result, the symmetry breaking pattern for the tri-component pairing is
$C_{4v}\times \mathbb{Z}^T_2 \rightarrow \mathbb{Z}_2$,
in which $\mathbb{Z}_2^T$ on the left side of the arrow represents $\langle T\rangle$, i.e., the group generated by the time reversal operation.
Since $|C_{4v}\times \mathbb{Z}_2^T|/| \mathbb{Z}_2|=8$,
there are eight degenerate solutions of  the  pairing configurations given by
\bea
s+i\eta_1 (e^{i\eta_2 \varphi/2}p_x+\eta_3 e^{-i\eta_2 \varphi/2}p_y),
\label{eq:8pairings}
\eea
in which $\eta_j=\pm 1$ ($j=1,2,3$).

On the other hand, the boundary condition deep in the $p$-wave bulk needs to be specified when minimizing the free energy, which amounts to fixing the chirality (i.e., $p_x+ip_y$ or $p_x-ip_y$) deep in the upper space.
The choice of the boundary condition explicitly breaks the time reversal and reflection symmetries since they both flip the chirality.
By putting the $s$- and chiral $p$-wave superconducting layers in contact with each other,
the junction structure further breaks the residual $C_4$ symmetry, \footnote{Here we note that strictly speaking, the residual symmetry group is not just $C_4$. The full residual symmetries of the $p_x+ip_y$ pairing are $\{ 1,r,r^2,r^3,TM_x,G^\prime TM_y,G TM_{x-y},G^{-1} TM_{x+y}\}$ which is isomorphic to $C_{4v}$,
where $r=G R(\hat{z},\pi/2)$;
$M_{f(x,y)}$ represents the spin-orbit coupled reflection with respect to the $f(x,y)=0$ plane;
$G$ is the gauge transformation by $\pi/4$;
and $G^\prime$ is the gauge transformation by $\pi/2$.
If we remove the time reversal operation, then the symmetry group becomes $C_4$.
}
where the action of the $C_4$ rotational operation on the chiral $p$-wave pairing is defined up to a gauge transformation.
The corresponding four degenerate tri-component pairing configurations among the eight ones in Eq. (\ref{eq:8pairings}) satisfy $\eta_2\eta_3=-\eta_c$,
when the boundary condition is chosen as $p_x+i\eta_c p_y$ where $\eta_c=\pm 1$.
Fig. \ref{tri} (a-d) display the configurations for the positive chirality case (i.e., $p_x+ip_y$),
and the other four negative chirality configurations can be obtained from those in  Fig. \ref{tri} by switching $\Delta_{px}$ and $\Delta_{py}$.

\section{Anisotropic magnetoelectric effect and edge magnetization}
\label{sec:Edge_S_response}

In this section, we discuss a novel type of anisotropic magnetoelectric effect in the tri-component pairing heterojunction.
Using a linear response approach, we show that a spatial variation of the electric potential can induce spin magnetizations along $z$-direction with a strength dependent on the direction of the electric field.
Since an edge corresponds to a change of the electric potential,
we conclude that the edge of the heterojunction carries  anisotropic spin magnetization if the potential change in the vicinity of the edge is slow enough such that the linear response approximation applies.
In the next section, we make a complimentary analysis on the opposite limit where the electric potential changes abruptly at the edge.
The anisotropic edge magnetization is shown to emerge  as the consequence of the splitting between the two branches of chiral Majorana edge modes.
Therefore, the ``soft" and ``hard" edge pictures on the edge magnetization are fully consistent with each other.

Before proceeding on,  we first note that there is no spin magnetization  along $z$-direction  for a uniform system.
This can be directly seen by noticing that in the tri-component pairing $s+i\eta_1 (e^{i\eta_2 \varphi/2}p_x+\eta_3 e^{-i\eta_2 \varphi/2}p_y)$, the Cooper pairings always occur  between up and down electrons,
thereby carry no spin angular momentum $S^z$.

\begin{figure}
\includegraphics[width=7cm]{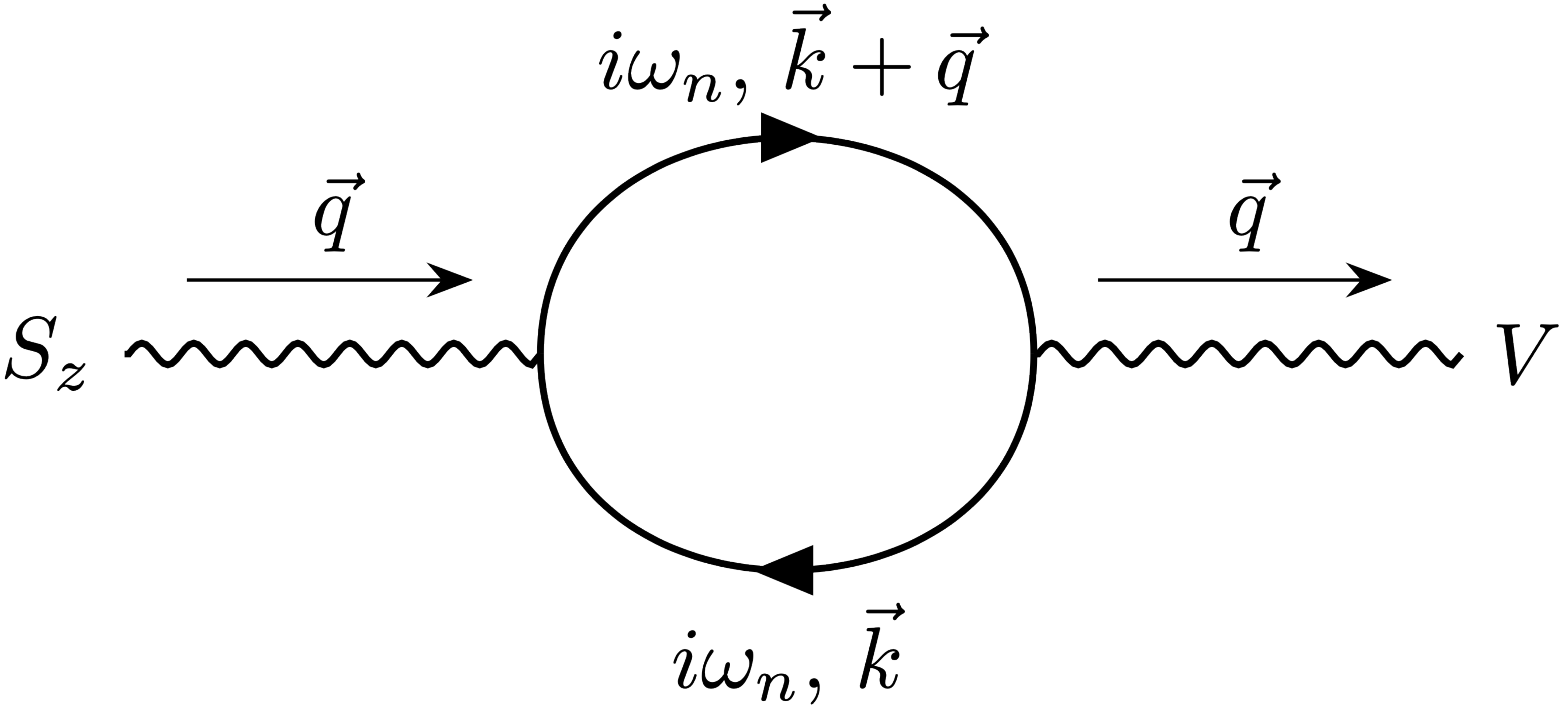}
\caption{
The Feynman diagram for the response of the spin magnetization 
$S^z$ to an external static electric potential $V$.
}
\label{linear}
\end{figure}

Next, we study the induced magnetization in the presence of a spatially varying electrical potential.
In the linear response theory, this is captured by the response of the spin magnetization density  $S^z(\vec{r})$ to an applied electric potential $V(\vec{r})$, as shown by the bubble diagram in Fig. \ref{linear}.
Assuming $V(\vec{r})$ to be slowly varying, we will only calculate the results up to linear order in the wavevector $\vec{q}$.
The solid lines in Fig. \ref{linear} represent the fermionic Green's
functions $G(i\omega_n,\vec{k})$ in the superconducting state where $\omega_n=(2n+1)\pi T$ ($n\in \mathbb{Z}$) is the fermionic Matsubara frequency,
and the dashed lines are the bosonic fields $S^z(\vec{r})$ or $V(\vec{r})$.
In the following, we assume that $\vec{r}$ represents the  two-dimensional spatial coordinates  within the junction interface.

In the momentum space within the BdG formalism, the pairing $\hat{\Delta}(\vec{k})$, the spin density $\hat{S}_z(\vec{q})$, and the particle number density $\hat{\rho}(\vec{q})$ can be represented as
\bea
\hat{S}^z(\vec{q})&=&\psi^\dag(\vec{k}+\vec{q}) S^z(\vec{q})\psi(\vec{k}), \nn\\
\hat{\rho}^x(\vec{q})&=&\psi^\dag(\vec{k}+\vec{q}) \rho(\vec{q})\psi(\vec{k}),\nn\\
\hat{\Delta}(\vec{k})&=&\psi^\dag(\vec{k}) \Delta(\vec{k})\psi^{\dagger,T}(\vec{k}),
\eea
in which
$
\psi(\vec{k}) = (c_\uparrow(\vec{k}),c_\downarrow(\vec{k}),c_\uparrow^\dag(-\vec{k}),
c_\downarrow^\dag(-\vec{k}))^T,
$
and the $4\times 4$ matrix kernels are
\bea
S^z(\vec{q})&=&\frac{1}{4} \sigma^z \tau^z,~~\rho(\vec{q})=\frac{1}{2} \tau^z,\nn\\
\Delta(\vec{k}) &=& -|\Delta_s| \sigma^y\tau^y-\frac{|\Delta_p|}{k_f}\big[(k_x+k_y)\sigma^x\tau^y \cos\frac{\varphi}{2}\nn\\
&&+(k_x-k_y)\sigma^x\tau^x\sin\frac{\varphi}{2}\big],
\eea
in which $\tau^j$ ($j=x,y,z$) are the Pauli matrices in the Nambu space,
and the tri-component structure $s+i (e^{i\frac{\varphi}{2}}p_x+ e^{-i \frac{\varphi}{2}}p_y)$ is assumed.
For simplicity, we take a rotationally invariant band dispersion $\xi(\vec{k})=\frac{\hbar^2}{2m}(k^2-k_f^2)$.
Using the Green's function
\bea
G(i\omega_n,\vec{k})=\frac{1}{i\omega_n-\xi(\vec{k})\tau^z-\Delta(\vec{k})},
\eea
the diagram in Fig. \ref{linear} can be evaluated as
\begin{eqnarray}
	\chi(\vec{q}) &=& -\int\frac{d^2k}{(2\pi)^2}\frac{1}{\beta}\sum_{i\omega_n} \text{Tr}[S_z G(i\omega_n,\vec{k}+\vec{q}) V
	G(i\omega_n,\vec{k})]\nonumber\\
	&=&\chi_0 (iq_x+iq_y),
	\label{eq:response}
\end{eqnarray}
in which within the limit $|\Delta_s|,|\Delta_p|\ll T$ (i.e., close to the superconducting transition temperature), $\chi_0$  is calculated to be
\bea
\chi_0\approx\frac{7\zeta(3)}{8\sqrt{2}\pi^2} N_F\frac{1}{T^2} \frac{|\Delta_p\Delta_s|}{k_f}\cos\frac{\varphi}{2},
\eea
where $\zeta$, $N_F$, and $T$ are the Riemann zeta function, the density of states at Fermi level, and the temperature, respectively.
In Eq. (\ref{eq:response}), the $\vec{q}$-independent terms vanish and only the terms linear in $\vec{q}$ are kept.
Detailed calculations are included in Appendix \ref{app:linear_response}.

The form of $\chi(\vec{q})$ in Eq. (\ref{eq:response}) implies the following response relation in real space,
\bea
	S^z (\vec{r}) =\chi_0 (\partial_x V+\partial_y V) = \sqrt{2}\chi_0 \hat{n}_0 \cdot \nabla V,
	\label{eq:response_real_space}
\eea
in which $\hat{n}_0=\frac{1}{\sqrt{2}}(1,1,0)$.
As is clear from Eq. (\ref{eq:response_real_space}),
the response is anisotropic since there is a special direction $\hat{n}_0$,
which is simply a consequence of the breaking of the $C_4$ symmetry.
Also notice that the two sides of Eq. (\ref{eq:response_real_space}) are both invariant under the unbroken symmetry transformation $TM_{x-y}$.
Indeed, the invariance under $TM_{x-y}$ is able to completely determine $\hat{n}_0$ to be parallel with the $(110)$-direction.

\begin{figure}
\includegraphics[width=6cm]{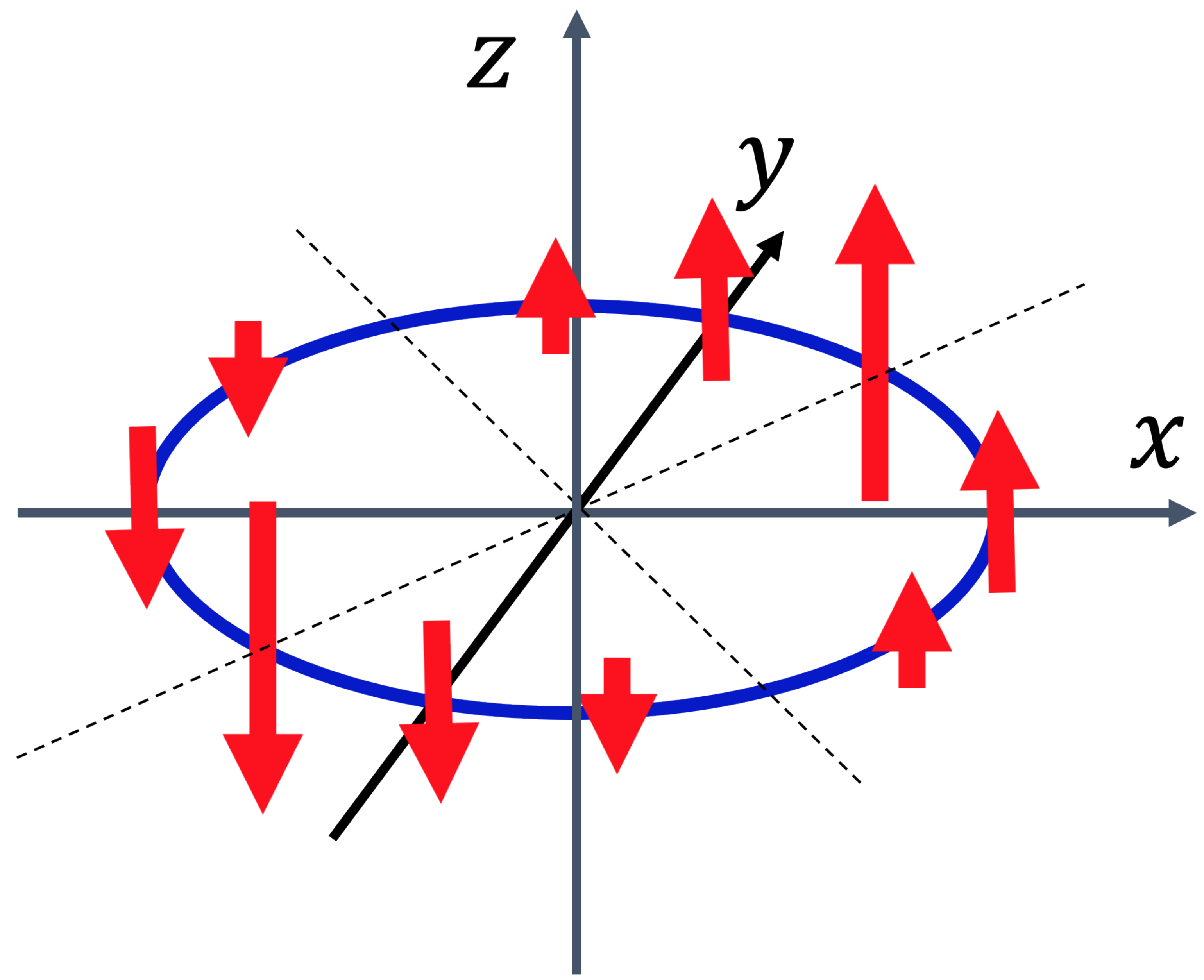}
\caption{
The anisotropic edge magnetization on a circular boundary of the junction.
The edge is represented by the blue circle.
The direction and magnitude of the edge magnetization are represented by 
the direction and height of the red arrows, respectively.
}
\label{illus}
\end{figure}

Finally we note that the edge can be modeled by a change of the electric potential.
The potential in the vacuum side is higher than the Fermi energy in the bulk so that the electrons in the vacuum are completely depleted.
Consider a ``soft" edge where the electric potential varies slowly.
Since $\nabla V=|\vec{\nabla}V|(\cos\theta, \sin\theta,0)$ is parallel to the normal direction of the edge,
it is clear from Eq. (\ref{eq:response_real_space}) that a spin magnetization emerges on the edge.
 For a rough estimation, $|\vec{\nabla}V|$ can be approximated as $\sim\epsilon_f/\xi_c$, where $\epsilon_f=\frac{\hbar^2}{2m}k_f^2$ is the Fermi energy and $\xi_c$ is the coherence length.
 Therefore the edge magnetization along the $z$-axis can be estimated as
\begin{equation}
    S_z(\theta) \sim \sqrt{2}\chi_0 \frac{\epsilon_f}{\xi_c} (\cos\theta
+ \sin\theta)\,,
\label{szt}
\end{equation}
 which is highly dependent on the  normal direction of the edge.
 Assuming the edge to be
in a circular shape,
 the edge magnetization along the $z$-direction is illustrated in Fig.~\ref{illus}, where the height of  the red arrows indicate the strength of the spin polarizations.

\section{Edge state picture of the edge magnetization}

In this section, we consider a ``hard" edge which is assumed to be an  infinite straight  line.
The system lies on one side of the edge, and the other side is the vacuum.
The boundary condition is taken such that the wavefunction vanishes at the edge and in the vaccum.
We show that the edge magnetization discussed in Sec. \ref{sec:Edge_S_response} with a ``soft" edge can alternatively be understood in the edge state picture.

For simplification of discussions, we perform a rotation of the coordinate  system defined as
\begin{equation}
	\left(
	\begin{matrix}
	x\\
	y
	\end{matrix}
	\right) =
	\left(
	\begin{matrix}
	\cos\theta && -\sin\theta\\
	\sin\theta && \cos\theta
	\end{matrix}
	\right)
	\left(
	\begin{matrix}
	x^\prime\\
	y^\prime
	\end{matrix}
	\right).
\end{equation}
In the rotated basis, $\hat{x}^\prime$ is along the normal direction $\hat{n}=\hat{x}\cos\theta+\hat{y}\sin\theta$ of the edge,
and $k_y^\prime$ is a good quantum number.
After the rotation, the superconducting pairing gap function is transformed into
\bea
\hat{\Delta}^\prime=\frac{1}{k_f}[\Delta_{px}^\prime(-i\partial^\prime_x)+\Delta_{py}^\prime k_y^\prime]\sigma^z i\sigma^y+\Delta_si\sigma^y,
\eea
in which
\bea
\Delta_{p\nu}^\prime&=&
|\Delta_p| [(\cos\theta+\sin\theta)\cos\frac{\varphi}{2}\nn\\
&&+\epsilon(\nu) i (\cos\theta-\sin\theta)\sin\frac{\varphi}{2}],
\eea
where $\nu=x,y$, and $-\epsilon(x)=\epsilon(y)=1$.
To further simplify the problem,
 a gauge transformation can be performed to absorb the phase of $\Delta^\prime_{px}$.
Then the pairing acquires the form
\bea
\hat{\Delta}^{\prime\prime}=\frac{1}{k_f}[\Delta_{px}^{\prime\prime}(-i\partial^\prime_x)+\Delta_{py}^{\prime\prime}k_y^\prime]\sigma^z i\sigma^y+\Delta_s^{\prime\prime}i\sigma^y,
\eea
in which
\bea
\Delta_{px}^{\prime\prime}&=&|\Delta_p|\sqrt{1+\sin(2\theta)\cos\varphi},\nn\\
\Delta_{py}^{\prime\prime}&=&|\Delta_p|\frac{\cos(2\theta)\cos\varphi+i\sin\varphi}{\sqrt{1+\sin(2\theta)\cos\varphi}},\\
\Delta_{s}^{\prime\prime}&=&i\Delta_s \frac{\cos\frac{\varphi}{2}(\cos\theta+\sin\theta)+i\sin\frac{\varphi}{2} (\cos\theta-\sin\theta) }{\sqrt{1+\sin(2\theta)\cos\varphi}}.\nn
\label{eq:2prime_pairing}
\eea
In what follows, we assume that the junction occupies the $x^\prime<0$ region, whereas $x^\prime>0$ is the vacuum.
The boundary condition is taken such that the wavefunction vanishes when $x^\prime\geq 0$.

\begin{figure}
\includegraphics[width=8.5cm]{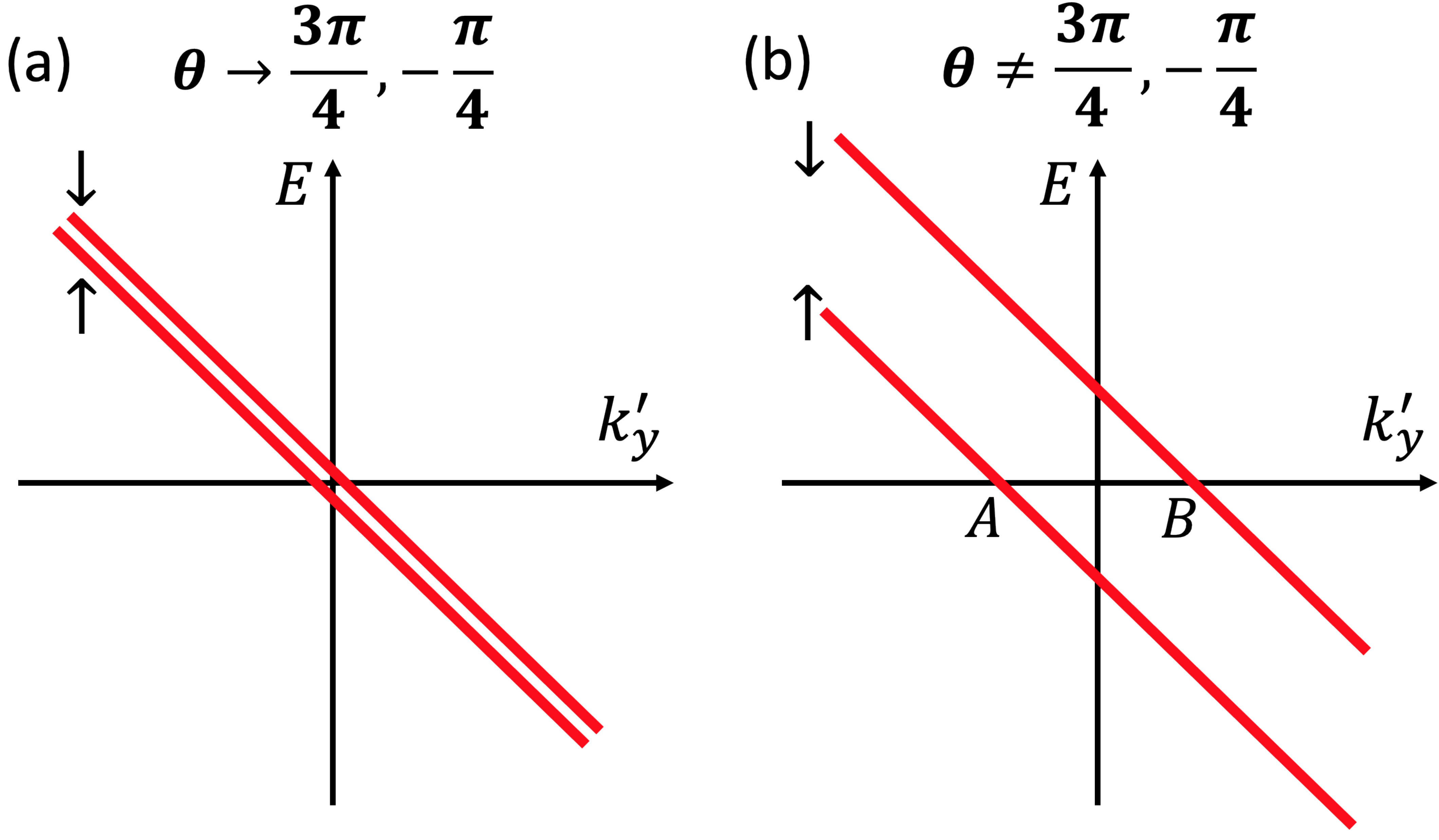}
\caption{Dispersions of the chiral edge Majorana modes for ($a$) $\theta=\frac{3\pi}{4},-\frac{\pi}{4}$ and ($b$) $\theta\neq\frac{3\pi}{4},-\frac{\pi}{4}$, 
where $\theta\in[-\pi,\pi]$.
}
\label{sp}
\end{figure}

The general solutions of the edge states are rather complicated.
To illustrate the essential physics, it is enough to consider the limit $|\Delta_s|\ll |\Delta_p|$.
The strategy is  first solving the edge states for $k_y^\prime=0$, and then a nonzero $k_y^\prime$ can be included using a $k\cdot p$ perturbation method.
In the absence of the $s$-wave component, there are two Majorana zero modes localized around the boundary for $k_y^\prime=0$.
In the weak pairing limit $|\Delta_p|\ll \epsilon_f$, the wavefunctions of the two zero modes can be solved as \cite{Yang2017}
\bea
\Phi_{\uparrow}(x^\prime) &=& ( e^{-i\frac{\pi}{4}} ,0 ,0 ,e^{i\frac{\pi}{4}} )^T u(x),\nn\\
\Phi_{\downarrow}(x^\prime) &=& ( 0,e^{-i\frac{\pi}{4}} ,e^{i\frac{\pi}{4}},0 )^T u(x),
\eea
in which $u(x)=\frac{1}{\sqrt{N}} \sin(k_f x) e^{\frac{m|\Delta_p|}{\hbar k_f}x}$,
where $N$ is a normalization factor.
Since $|\Delta_s|\ll |\Delta_p|$, the $s$-wave pairing can be treated using a first order perturbation.
It is straightforward to verify that the projection of $\hat{\Delta}_s$ (defined in Eq. (\ref{eq:Delta_pairing})) to the basis $\{\Phi_\uparrow,\Phi_\downarrow\}$ is $-(\text{Im}\Delta_s^{\prime\prime}) s^z$,
where $s^\alpha$ ($\alpha=x,y,z$) are the Pauli matrices in the space spanned by $\{\Phi_\uparrow,\Phi_\downarrow\}$, and $\text{Im}\Delta_s^{\prime\prime}$ can be read from Eq. (\ref{eq:2prime_pairing}).
Therefore, while the Majorana modes remain at zero energy under the real part of $\Delta_s^{\prime\prime}$, the imaginary part of $\Delta_s^{\prime\prime}$ opens a gap on the edge.

Next we move to a nonzero $k_y^\prime$.
The $k\cdot p$ Hamiltonian can be obtained by projecting the pairing along the $y^\prime$-direction to the basis $\{\Phi_\uparrow,\Phi_\downarrow\}$,
and the result is $-\frac{\text{Im} \Delta_{py}^{\prime\prime}}{k_f}k_y^\prime s^0$
where $s^0$ is the $2\times 2$ identity matrix.
Combining with the contribution from the $\text{Im}\Delta_s^{\prime\prime}$ term,
the dispersions of the two  chiral Majorana edge fermions  can be derived as
\bea
E_{\eta}(k_y^\prime)=-\frac{\text{Im} \Delta_{py}^{\prime\prime}}{k_f}k_y^\prime-\eta\text{Im}\Delta_s^{\prime\prime},
\label{eq:edge_disperse}
\eea
in which $E_\eta(k_y^\prime)$ is the dispersion of the $\eta$-branch of the chiral modes,
where $\eta=1$ ($-1$) for $\uparrow$ ($\downarrow$).
Therefore, when an $s$-wave component  is present in the pairing,
the two edge modes split by an energy gap $\Delta E = 2
{\rm{Im}}\Delta_s^{\prime\prime}$.
Since ${\rm{Im}}\Delta_s^{\prime\prime}$ vanishes when $\theta=3\pi/4,-\pi/4$,
the spin up and down chiral branches coincide with each other as shown in Fig. \ref{sp}.
When $\theta\neq3\pi/4,-\pi/4$, the two branches split due to the opening of the gap as shown in Fig. \ref{sp} (b).

The two branches of chiral Majorana edge modes are spin polarized.
As can be seen from Eq. (\ref{eq:edge_disperse}), within the approximation of a linear dispersion, the occupation range of $k_y^\prime$ for the $\lambda$-branch of the chiral mode is
$\epsilon_\lambda  \frac{\text{Im}\Delta_s^{\prime\prime} }{\text{Im} \Delta_{py}^{\prime\prime}} k_f\leq k_y^\prime\leq k_f$, in which $\epsilon_\lambda=1$ ($-1$) for $\lambda=\uparrow$ ($\downarrow$).
This leads to an imbalance in the occupation range between the up and down chiral edge modes corresponding to the line segment between the points $A$ and $B$ in Fig. \ref{sp} (b).
As a consequence,  a spin polarization develops   on the edge,
which has a direction-dependence proportional to ${\rm{Im}}{\Delta_s^{\prime\prime}}/{\rm{Im}}{\Delta_y^{\prime\prime}} \sim(\sin\theta+\cos\theta)$.
In particular, this result is consistent with what have been obtained in Sec. \ref{sec:Edge_S_response} as shown in Fig. \ref{illus}.
Thus we see that the ``soft" and ``hard" edge pictures on the edge magnetization are fully consistent with each other.

Finally we also note that experiments on the anisotropic effect of the edge magnetization in the heterojunction could be potentially useful for testing 
the  gap function symmetries of unconventional superconductors.

\section{Conclusion}

In conclusion, we have studied the heterjunction with one side possessing 
the chiral $p$-wave (i.e., $p_x\pm ip_y$) and the other side the conventional 
$s$-wave pairing gap functions, respectively.
By employing a Ginzburg-Landau free energy analysis, the pairing gap function 
in the junction region is shown to exhibit a frustrated tri-component 
structure as $s+ i\eta_1 (e^{ i\eta_2 \varphi/2}p_x+\eta_3 e^{- i\eta_2 \varphi/2}p_y)$,
where $\varphi$ is the phase difference between the $p_x$ and $p_y$ 
components, and $\eta_j=\pm 1$ ($j=1,2,3$).
By solving the chiral Majorana  edge modes with the tri-component pairing, 
we find that the edge of the junction carries an anisotropic spin magnetization,
where the anisotropy originates from the breaking of the rotational symmetry.
In addition, the edge magnetization is consistent with a novel type of 
anisotropic magnetoelectric effect, which is analyzed through the 
linear response calculation.


\begin{widetext}
\appendix

\section{The Ginzburg-Landau free energy}
\label{app:GL}

For simplicity, we will consider a system with isotropic Fermi surface.
As a result, the $\beta_p^\prime$ term vanishes.
Only keeping the spatially uniform parts, the  free energy up to quartic orders is
\begin{flalign}
	&f_{spp}=\alpha_s|\Delta_s|^2+\alpha_p(|\Delta_{px}|^2+
	|\Delta_{py}|^2) + \beta_s |\Delta_s|^4 +
	\beta_p (|\Delta_{px}|^4+|\Delta_{py}|^4) +g_{pp} [(\Delta_{px}^*\Delta_{py})^2 + (\Delta_{py}^*\Delta_{px})^2]
	 \nonumber \\
	&~~~+\nu_p |\Delta_{px}|^2|\Delta_{py}|^2+\gamma_1 (|\Delta_{px}|^2+|\Delta_{py}|^2)|\Delta_s|^2
	+g_{sp} [\Delta_s^{*2}(\Delta_{px}^2+\Delta_{py}^2) +
	\Delta_s^{2}(\Delta_{px}^{*2}+\Delta_{py}^{*2})].
\label{eq:free_energy_app}
\end{flalign}
While the coefficients of the quadratic terms depend on the interactions which rely on the details of the pairing mechanism,
 the coefficients of the quartic terms are not dependent on the interaction strength within a tree-level approximation and can be determined from the diagrams in Fig. \ref{fig:GL_diagrams},
in which the the superconducting order parameters are given by
\begin{eqnarray}
\hat{\Delta}_s&=&\frac{\Delta_s}{2}\sum_k c_k^{\dag}i\sigma^y (c_{-k}^\dag)^T,\nn\\
\hat{\Delta}_{px}&=&\frac{\Delta_{px}}{2k_f} \sum_kc_k^{\dag}(k_x \sigma^z)i\sigma^y (c_{-k}^{\dag})^T,\nn\\
\hat{\Delta}_{py}&=&\frac{\Delta_{py}}{2k_f} \sum_k c_k^{\dag}(k_y \sigma^z)i\sigma^y (c_{-k}^{\dag})^T,
\end{eqnarray}
where  $c^\dagger_k=(c^\dagger_{k\uparrow}~c^\dagger_{k\downarrow})$ is a two-component row vector.

\begin{figure*}
\includegraphics[width=15cm]{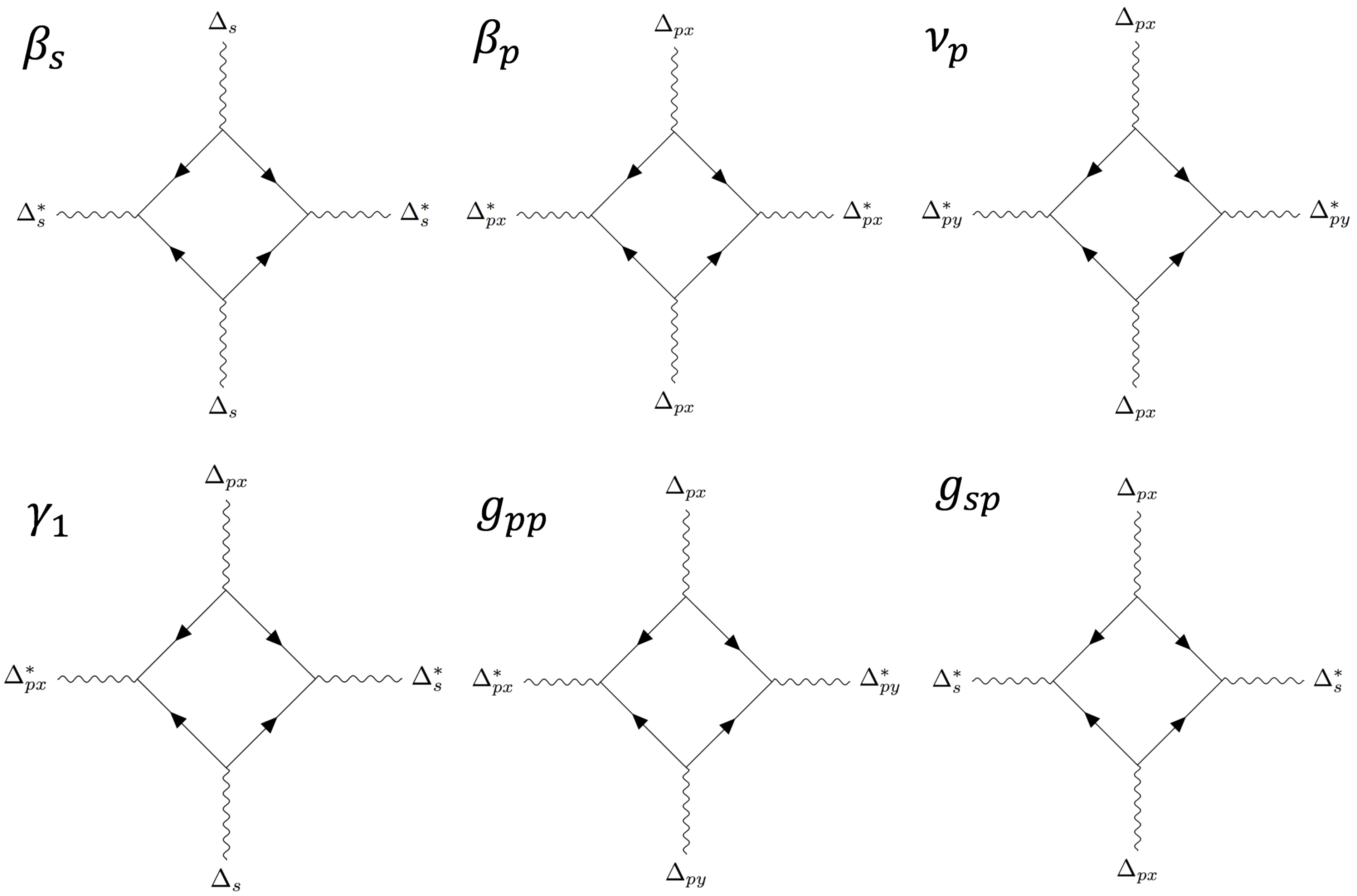}
\caption{Diagrams determining the coefficients in the Ginzburg-Landau free energy.
}
\label{fig:GL_diagrams}
\end{figure*}

Keeping only the static and uniform terms (i.e., zero frequency and zero momentum), we obtain
\begin{eqnarray}
\beta_s &=&\frac{3}{2} \hat{\beta}_0  {\rm{Tr}}\{(i\sigma^y)^\dag(i\sigma^y)(i\sigma^y)^\dag(i\sigma^y)\} ,\nn\\
\beta_{p} &=& \frac{3}{2}\hat{\beta}_0\frac{1}{k_f^4}{\rm{Tr}}\{(i k_\alpha\sigma^z\sigma^y)^\dag(i k_\alpha\sigma^z\sigma^y)(i k_\alpha\sigma^z\sigma^y)^\dag(i k_\alpha\sigma^z\sigma^y),\nn\\
\nu_{p} &=&6 \hat{\beta}_0\frac{1}{k_f^4}{\rm{Tr}}\{(i k_x\sigma^z\sigma^y)^\dag(i k_x\sigma^z\sigma^y)(i k_y\sigma^z\sigma^y)^\dag(i k_y\sigma^z\sigma^y),\nn\\
\gamma_1 &=& 6\hat{\beta}_0 \frac{1}{k_f^2}{\rm{Tr}}\{(i k_\alpha\sigma^z\sigma^y)^\dag (i k_\alpha \sigma^z\sigma^y)(i\sigma^y)^\dag(i\sigma^y)\}, \nn\\
g_{pp} &=&\frac{3}{2}\hat{\beta}_0\frac{1}{k_f^4}{\rm{Tr}}\{(i k_x\sigma^z\sigma^y)^\dag(i k_y\sigma^z\sigma^y)(i k_x\sigma^z\sigma^y)^\dag(i k_y\sigma^z\sigma^y)\},\nn\\
g_{sp} &=& 6 \hat{\beta}_0\frac{1}{k_f^2} {\rm{Tr}}\{(i\sigma^y)^\dag(i k_\alpha\sigma^z\sigma^y) (i\sigma^y)^\dag(i k_\alpha \sigma^z\sigma^y))\},
\label{eq:coeffs}
\end{eqnarray}
in which $k_\alpha$ can be taken as either $k_x$ or $k_y$, and the operation $\hat{\beta_0}$ acting on the expression to the right of it is defined as
\begin{eqnarray}
\hat{\beta}_0[\cdot\cdot\cdot]=\frac{1}{\beta}\frac{1}{L^3}\sum_{\omega_m,k}\frac{1}{(\omega_m^2 + \xi_k^2)^2} [\cdot\cdot\cdot],
\end{eqnarray}
where $\xi_k=\hbar^2k^2/2m-\epsilon_F$, and $L^3$ is the volume of the system.
In the weak pairing limit, a linearization of the dispersion can be performed.
Changing the integration over $\vec{k}$ to spherical coordinates, we have
\bea
\hat{\beta}_0[\cdot\cdot\cdot] =N_F\frac{1}{\beta}\sum_n \int_{-\infty}^{\infty} d\epsilon \int_0^\pi \sin\theta d\theta \int_0^{2\pi} d\phi \frac{1}{\big([(2n+1)\pi/\beta]^2+\epsilon^2\big)^2}[\cdot\cdot\cdot],
\label{eq:beta_0_linearize}
\eea
in which $N_F$ is the density of states at Fermi energy.

Plugging Eq. (\ref{eq:beta_0_linearize}) into Eq. (\ref{eq:coeffs}), performing the integrations, and summing over the Matsubara frequencies, we obtain
\bea
\beta_s =\frac{3}{2}\beta,~
\beta_{p}=\frac{3}{10}\beta,~
\nu_p=\frac{2}{5}\beta,~
\gamma_1 =2\beta,~
g_{pp} =\frac{1}{10}\beta,~
g_{sp} =2\beta.
\eea
in which
\bea
\beta=\frac{7\zeta(3)N_F}{8\pi^2T^2}.
\eea
Notice that since $2(\beta_p-g_{pp})=\nu_p$, the $p$-wave terms in Eq. (\ref{eq:free_energy_app}) can be recombined into the form in Eq. (\ref{eq:f_1}).

Finally we note that the coefficients determined in this section are not accurate in real situations, since there are notable renormalization effects, particularly when $T$ is close to $T_c$.

\section{The linear response of the anisotropic magnetoelectric effect}
\label{app:linear_response}

We work in the ordered phase and calculate the correlation function between $S^z$ and $\rho$.
In the following calculations, we take the pairing as $-is+e^{i\varphi/2}p_x+e^{-i\varphi/2}p_y$.
The pairing is taken as
\begin{flalign}
&\frac{\Delta_p}{k_f} (e^{i\varphi/2}k_x+e^{-i\varphi/2}k_y) \sigma^zi\sigma^y-i\Delta_si\sigma^y\nn\\
&=\left(\begin{array}{cc}
0&\frac{\Delta_p}{k_f}\cos(\frac{\varphi}{2})(k_x+k_y)+i\big[\frac{\Delta_p}{k_f} \sin(\frac{\varphi}{2})(k_x-k_y)-\Delta_s) \big] \\
\frac{\Delta_p}{k_f}\cos(\frac{\varphi}{2})(k_x+k_y)+i\big[\frac{\Delta_p}{k_f} \sin(\frac{\varphi}{2})(k_x-k_y)+\Delta_s) \big]&0
\end{array}\right),
\end{flalign}
in which both $\Delta_p$ and $\Delta_s$ are real and positive.
In the spin up sector, the BdG Hamiltonian is of the form
\bea
H_{\uparrow} (\vec{k})&=&\left(\begin{array}{cc}
\xi(\vec{k})& \frac{\Delta_p}{k_f}\cos(\frac{\varphi}{2})(k_x+k_y)-i\big[ \frac{\Delta_p}{k_f} \sin(\frac{\varphi}{2})(-k_x+k_y)+\Delta_s) \big]\\
\frac{\Delta_p}{k_f}\cos(\frac{\varphi}{2})(k_x+k_y)+i\big[ \frac{\Delta_p}{k_f} \sin(\frac{\varphi}{2})(-k_x+k_y)+\Delta_s) \big]& -\xi(-\vec{k})
\end{array}\right)\nn\\
&=&\xi(\vec{k})\iota^z+\frac{\Delta_p}{k_f} \cos(\frac{\varphi}{2}) (k_x+k_y) \iota^x
+\big[ \frac{\Delta_p}{k_f}\sin(\frac{\varphi}{2} )(-k_x+k_y))+\Delta_s \big]\iota^y.
\eea
Since the spin up and down sectors are related by a particle-hole transformation,
it is enough to work in the spin up sector.
We also note that the matrix kernels for $S^z$ and $\rho$ in the spin up sector are $\iota^z$ and $\frac{1}{2}\iota^0$, respectively, where $\iota^0$ is the $2\times 2$ identity matrix.
In what follows, we write $\iota^\alpha$ as $\sigma^\alpha$ ($\alpha=0,x,y,z$) for simplicity.

In the imaginary time formalism with, the diagram in Fig. \ref{linear} can be evaluated as
\bea
\chi(\vec{q})&=&-\int\frac{d^2\vec{k}}{(2\pi)^2}\frac{1}{\beta} \sum_{i\omega_n} \text{tr}
\big[
\frac{1}{2}\sigma^0 \frac{1}{i\omega_n-H_{\uparrow}(\vec{k}+\vec{q})}
\sigma^3 \frac{1}{i\Omega_n-H_{\uparrow}(\vec{k})}
\big]\nn\\
&=&-\int\frac{d^2\vec{k}}{(2\pi)^2}\frac{1}{\beta} \sum_{i\omega_n}
\frac{1}{\omega_n^2+\xi^2(\vec{k}+\vec{q})+\frac{\Delta_p^2}{k_f^2} \cos^2(\frac{\varphi}{2})(k_x+q_x+k_y+q_y)^2+\big[ \frac{\Delta_p}{k_f}\sin(\frac{\varphi}{2})(-k_x-q_x+k_y+q_y)+\Delta_s  \big]^2 }\nn\\
&&\times
\frac{1}{\omega_n^2+\xi^2(\vec{k})+\frac{\Delta_p^2}{k_f^2} \cos^2(\frac{\varphi}{2})(k_x+k_y)^2+\big[ \frac{\Delta_p}{k_f}\sin(\frac{\varphi}{2})(-k_x+k_y)+\Delta_s  \big]^2 }\nn\\
&&\times\text{tr} \big[
\frac{\sigma^0}{2} \big( i\omega_n+\xi(\vec{k}+\vec{q}) \sigma^z+\frac{\Delta_p}{k_f} \cos(\frac{\varphi}{2})(k_x+q_x+k_y+q_y) \sigma^x+[\frac{\Delta_p}{k_f}\sin(\frac{\varphi}{2}) (-k_x-q_x+k_y+q_y)+\Delta_s ]\sigma^y  \big)\nn\\
&&~~~\cdot \sigma^z \big( i\omega_n+\xi(\vec{k}) +\frac{\Delta_p}{k_f} \cos(\frac{\varphi}{2}) (k_x+k_y) \sigma^x+[\frac{\Delta_p}{k_f} \sin(\frac{\varphi}{2})(-k_x+k_y)+\Delta_s  ]\sigma^y  \big)
\big].
\label{eq:response_chi}
\eea
The trace term in Eq. (\ref{eq:response_chi}) can be evaluated to be
\bea
\text{tr}[\cdot\cdot\cdot]=-i\frac{\Delta_p}{k_f} \cos(\frac{\varphi}{2})
\big[  q_x(\frac{2\Delta_p}{k_f}\sin(\frac{\varphi}{2}) k_y+\Delta_s)+q_y
(-\frac{2\Delta_p}{k_f}\sin(\frac{\varphi}{2}) k_x+\Delta_s) \big],
\eea
in which the linear in $\omega_n$ terms are neglected since they sum to zero after Matsubara frequency summation.
Since the numerator  of Eq. (\ref{eq:response_chi}) is already linear in $\vec{q}$,
the $\vec{q}$'s in the denominator can be set to be zero since we only need the results up to $O(\vec{q})$.
Then we arrive at
\bea
\chi(\vec{q})=iq_x \chi_x+iq_y\chi_y,
\eea
in which
\bea
\chi_x&=& \frac{\Delta_p}{k_f} \cos(\frac{\varphi}{2}) \int \frac{d^2\vec{k}}{(2\pi)^2}
\frac{1}{\beta} \sum_{i\omega_n}
\frac{\frac{2\Delta_p}{k_f}\sin(\frac{\varphi}{2})k_y+\Delta_s }{\omega_n^2+\xi^2(\vec{k})+\frac{\Delta_p^2}{k_f^2} \cos^2(\frac{\varphi}{2})(k_x+k_y)^2+\big[ \frac{\Delta_p}{k_f}\sin(\frac{\varphi}{2})(-k_x+k_y)+\Delta_s  \big]^2 },\nn\\
\chi_y&=& \frac{\Delta_p}{k_f} \cos(\frac{\varphi}{2}) \int \frac{d^2\vec{k}}{(2\pi)^2}
\frac{1}{\beta} \sum_{i\omega_n}
\frac{-\frac{2\Delta_p}{k_f}\sin(\frac{\varphi}{2})k_x+\Delta_s }{\omega_n^2+\xi^2(\vec{k})+\frac{\Delta_p^2}{k_f^2} \cos^2(\frac{\varphi}{2})(k_x+k_y)^2+\big[ \frac{\Delta_p}{k_f}\sin(\frac{\varphi}{2})(-k_x+k_y)+\Delta_s  \big]^2 }.
\eea

Next to simplify the expressions of $\chi_x$ and $\chi_y$, we perform a change of variable
\bea
k_x^\prime=\frac{1}{\sqrt{2}}(k_x+k_y), ~k_y^\prime=\frac{1}{\sqrt{2}}(-k_x+k_y).
\eea
Then we have
\bea
\chi_x=A_x+A_y+A_s,~\chi_y=-A_x+A_y+A_s,
\eea
in which
\bea
A_\alpha&=& \frac{\Delta_p}{k_f} \cos(\frac{\varphi}{2}) \int \frac{d^2\vec{k}^\prime}{(2\pi)^2}
\frac{1}{\beta} \sum_{i\omega_n}
\frac{\frac{\sqrt{2}\Delta_p}{k_f}\sin(\frac{\varphi}{2})k_\alpha^\prime }{\omega_n^2+\xi^2(\vec{k}^\prime)+\frac{2\Delta_p^2}{k_f^2} \cos^2(\frac{\varphi}{2})k_x^{\prime 2}+\big[ \frac{\sqrt{2}\Delta_p}{k_f}\sin(\frac{\varphi}{2})k_y^\prime+\Delta_s  \big]^2 },\nn\\
A_s&=&\frac{\Delta_p}{k_f} \cos(\frac{\varphi}{2}) \int \frac{d^2\vec{k}^\prime}{(2\pi)^2}
\frac{1}{\beta} \sum_{i\omega_n}
\frac{\Delta_s }{\omega_n^2+\xi^2(\vec{k}^\prime)+\frac{2\Delta_p^2}{k_f^2} \cos^2(\frac{\varphi}{2})k_x^{\prime 2}+\big[ \frac{\sqrt{2}\Delta_p}{k_f}\sin(\frac{\varphi}{2})k_y^\prime+\Delta_s  \big]^2 },
\eea
in which $\alpha=x,y$.
Clearly, $A_\alpha$ ($\alpha=x,y$) vanishes since the numerator is odd under the integration over $\int dk_\alpha^\prime$.

In the limit $\Delta_s,\Delta_p\ll T$, the dependence on the order parameters in the denominators of $A_s$ can be neglected, and we have
\bea
\chi_x=\chi_y\approx \frac{\Delta_p\Delta_s}{\sqrt{2}k_f} \cos(\frac{\varphi}{2})
\int \frac{d^2\vec{k}}{(2\pi)^2} \frac{1}{\beta}\sum_{i\omega_n} \frac{1}{(\omega_n^2+\xi^2(\vec{k}))^2}.
\eea
The integral can be evaluated as
\bea
\int \frac{d^2\vec{k}}{(2\pi)^2} \frac{1}{\beta}\sum_{i\omega_n} \frac{1}{(\omega_n^2+\xi^2(\vec{k}))^2}=N_0 \frac{1}{\beta} \sum_{i\omega_n} \int d\epsilon \frac{1}{(\omega_n^2+\epsilon^2)^2}
=N_0\frac{1}{\beta} \sum_{n\in\mathbb{Z}} \frac{\pi}{2}\frac{1}{|2\pi n/T|^3}
=\frac{7\zeta(3)}{8\pi^2} N_0 \frac{1}{T^2},
\eea
in which $\zeta$ is the Riemann zeta function.

In summary, in the limit $\Delta_s,\Delta_p\ll T$, the response is
\bea
S^z=\chi_0 (\partial_x V+\partial_y V),
\eea
in which
\bea
\chi_0=\frac{7\zeta(3)}{8\sqrt{2}\pi^2} N_0\frac{1}{T^2} \frac{\Delta_p\Delta_s}{k_f}\cos(\frac{\varphi}{2}).
\eea

\end{widetext}

\begin{thebibliography}{10}

\bibitem{Kallin2016}
C. Kallin and J. Berlinsky, Rep. Prog. Phys. {\bf 79}, 054502 (2016).

\bibitem{Read2000}
N. Read and D. Green, Phys. Rev. B {\bf 61}, 10267 (2000).

\bibitem{DasSarma2006}
S. Das Sarma, C. Nayak, and S. Tewari, Phys. Rev. B {\bf 73}, 220502 (2006).

\bibitem{Fu2008}
L. Fu and C. L. Kane, Phys. Rev. Lett. {\bf 100}, 096407 (2008).

\bibitem{Sau2010}
J. D. Sau, R. M. Lutchyn, S. Tewari, and S. Das Sarma, Phys. Rev. Lett. {\bf 104}, 040502 (2010).

\bibitem{Teo2010}
J. ~C. ~Y. Teo and C.~ L. ~Kane, Phys. Rev. Lett. {\bf 104}, 046401 (2010).

\bibitem{Kitaev2003}
A. Kitaev, Annals of Physics {\bf 303}, 2 (2003).

\bibitem{Kitaev2006}
A. Kitaev, Annals of Physics {\bf 321}, 2 (2006).

\bibitem{Stone2006}
M. Stone and S.-B. Chung, Phys. Rev. B {\bf 73}, 014505 (2006).

\bibitem{Alicea2011}
J. Alicea, Y. Oreg, G. Refael, F. von Oppen, and M. ~P.~ A. Fisher, Nat. Phys. {\bf 7}, 412 (2011).

\bibitem{Halperin2012}
B. I. Halperin, Y. Oreg, A. Stern, G. Refael, J. Alicea, and F. von Oppen, Phys. Rev. B {\bf 85}, 144501 (2012).

\bibitem{Yu2010}
R. Yu, W. Zhang, H.-J. Zhang, S.-C. Zhang, X. Dai, and Z. Fang, Science {\bf 329}, 61 (2010).

\bibitem{Qi2010}
X.-L. Qi, T. L. Hughes, and S.-C. Zhang, Phys. Rev. B {\bf 82}, 184516 (2010).

\bibitem{Chung2011}
S. B. Chung, X.-L. Qi, J. Maciejko, and S.-C. Zhang, Phys. Rev. B {\bf 83}, 100512 (2011).

\bibitem{Mackenzie2003}
A. P. Mackenzie and Y. Maeno, Rev. Mod. Phys. {\bf 75}, 657 (2003).

\bibitem{Maeno2012}
Y. Maeno, S. Kittaka, T. Nomura, S. Yonezawa, and K. Ishida, J. Phys. Soc. Jpn. {\bf 81}, 011009 (2012).

\bibitem{Liu2015}
Y. Liu and Z.-Q. Mao, Physica C {\bf 514}, 339 (2015).

\bibitem{Maeno1994}
Y. Maeno, H. Hashimoto, K. Yoshida, S. Nishizaki, T. Fujita, J. G. Bednorz, and F. Lichtenberg, Nature (London) {\bf 372}, 532 (1994).

\bibitem{Joynt2002}
R. Joynt and L. Tallifer, Rev. Mod. Phys. {\bf 74}, 235 (2002).

\bibitem{Schemm2004}
E. R. Schemm, W. J. Gannon, C. M. Wishne, W. P.
Halperin, and A. Kapitulnik, Science {\bf 345}, 190 (2014).

\bibitem{Strand2009}
J. D. Strand, D. J. Van Harlingen, J. Kycia,   and W. P.  Halperin,  Phys. Rev. Lett. {\bf 103}, 59 (2009).

\bibitem{Avers2020}
K. E. Avers, W. J. Gannon, S. J. Kuhn, W. P. Halperin, J. A. Sauls, L. DeBeer-Schmitt, C. D. Dewhurst, J. Gavilano, G. Nagy, U. Gasser, and M. R. Eskildsen,
Nat. Phys. {\bf 16}, 531 (2020).

\bibitem{Kallin2009}
C. Kallin and A. J. Berlinsky, J. Phys.: Condens. Matter {\bf 21}, 164210
(2009)

\bibitem{Kallin2012}
C. Kallin, Rep. Prog. Phys. {\bf 75}, 042501 (2012).

\bibitem{Mackenzie2017}
A. P. Mackenzie, T. Scaffidi, C. W. Hicks, and Y. Maeno, npj Quantum Mater. {\bf 2},
40 (2017).

\bibitem{Ishida1998}
K. Ishida, H. Mukuda, Y. Kitaoka, K. Asayama, Z. Q. Mao,
Y. Mori, and Y. Maeno, Nature (London) {\bf 396}, 658 (1998).

\bibitem{Duffy2000}
J. A. Duffy, S. M. Hayden, Y. Maeno, Z. Mao, J. Kulda, and G. J. McIntyre, Phys. Rev. Lett. {\bf 85}, 5412 (2000).

\bibitem{Laube2000}
F. Laube, G. Goll, H. v. L\"ohneysen, M. Fogelstr\"om, and F. Lichtenberg, Phys. Rev. Lett. {\bf 84}, 1595 (2000).

\bibitem{Mackenzie1998}
A. P. Mackenzie, R. K. W. Haselwimmer, A. W. Tyler, G. G. Lonzarich, Y. Mori, S. Nishizaki, and Y. Maeno, Phys. Rev. Lett. {\bf 80}, 161 (1998).

\bibitem{Luke1998}
G. M. Luke, Y. Fudamoto, K. M. Kojima, M. I. Larkin, J. Merrin, B. Nachumi, Y. J. Uemura, Y. Maeno, Z. Q. Mao, Y. Mori, H. Nakamura and M. Sigrist,
Nature (London) {\bf 394}, 558 (1998).

\bibitem{Nelson2004}
K. Nelson, Z. Mao, Y. Maeno, and Y. Liu, Science {\bf 306}, 1151 (2004).

\bibitem{Xia2006}
J. Xia, Y. Maeno, P. T. Beyersdorf, M. M. Fejer, and A. Kapitulnik, Phys. Rev. Lett. {\bf 97}, 167002 (2006).

\bibitem{Kidwingira2006}
F. Kidwingira, J. Strand, D. Van Harlingen, and Y. Maeno, Science {\bf 314}, 1267 (2006).

\bibitem{Pustogow2019}
A. Pustogow, Y. Luo, A. Chronister, Y.-S. Su, D. A. Sokolov, F. Jerzembeck, A. P. Mackenzie, C. W. Hicks, N. Kikugawa, S. Raghu, E. D. Bauer, and S. E. Brown,
Nature {\bf 574}, 72 (2019).

\bibitem{Volovik1988}
G. Volovik, Phys. Lett. A {\bf 128}, 277 (1988)

\bibitem{Volovik1989}
G. Volovik and V. Yakovenko, J. Phys. Condens. Matter {\bf 1}, 5263 (1989)

\bibitem{Ivanov2001}
D. A. Ivanov, Phys. Rev. Lett. {\bf 86}, 268 (2001).

\bibitem{Kopnin1991}
N. B. Kopnin and M. M. Salomaa, Phys. Rev. B {\bf 44}, 9667 (1991).

\bibitem{Tewari2007}
S. Tewari, S. Das Sarma, C. Nayak, C. Zhang, and P. Zoller, Phys. Rev. Lett. {\bf 98}, 010506 (2007).

\bibitem{Chuanwei2008}
C. Zhang, S. Tewari, R. M. Lutchyn, and S. Das Sarma, Phys. Rev. Lett. {\bf 101}, 160401 (2008).

\bibitem{Cheng2010}
M. Cheng, K. Sun, V. Galitski, and S. Das Sarma, Phys. Rev. B {\bf 81}, 024504 (2010).

\bibitem{Qi2009}
X.-L. Qi, T. L. Hughes, S. Raghu, and S.-C. Zhang, Phys. Rev. Lett. {\bf 102}, 187001 (2009).






\bibitem{Laughlin1998}
R. B. Laughlin, Phys. Rev. Lett. {\bf 80}, 5188 (1998).

\bibitem{Senthil1999}
T. Senthil, J. B. Marston, and M. P. A. Fisher, Phys. Rev. B {\bf 60}, 4245 (1999).

\bibitem{Horovitz2003}
B. Horovitz and A. Golub, Phys. Rev. B {\bf 68}, 214503 (2003).

\bibitem{Hu2008}
Y. Jiang, D.-X. Yao, E. W. Carlson, H.-D. Chen, and J. Hu, Phys. Rev. B {\bf 77}, 235420 (2008).

\bibitem{Sato2010}
M. Sato, Y. Takahashi, and S. Fujimoto, Phys. Rev. B {\bf 82}, 134521 (2010).

\bibitem{Black2012}
A. M. Black-Schaffer, Phys. Rev. Lett. {\bf 109}, 197001 (2012).

\bibitem{Chubukov2012}
R. Nandkishore, L. S. Levitov, and A. V. Chubukov, Nat. Phys. 8, {\bf 158} (2012).

\bibitem{Wang2012}
W.-S. Wang, Y.-Y. Xiang, Q.-H. Wang, F. Wang, F. Yang, and D.-H. Lee, Phys. Rev. B {\bf 85}, 035414 (2012).

\bibitem{Kiesel2013}
M. L. Kiesel, C. Platt, W. Hanke, and R. Thomale, Phys. Rev. Lett. {\bf 111}, 097001 (2013).

\bibitem{Liu2013}
F. Liu, C.-C. Liu, K. Wu, F. Yang, and Y. Yao, Phys. Rev. Lett. {\bf 111}, 066804 (2013).

\bibitem{Black2014}
A. M. Black-Schaffer and C. Honerkamp, J. Phys.: Condens. Matter. {\bf26} 423201 (2014).

\bibitem{Liu2018}
C.-C. Liu, L.-D. Zhang, W.-Q. Chen, and F. Yang, Phys. Rev. Lett. {\bf 121}, 217001 (2018).

\bibitem{Kennes2018}
D. M. Kennes, J. Lischner, and C. Karrasch, Phys. Rev. B {\bf 98}, 241407(R) (2018).

\bibitem{Yang2018}
Z. Yang, S. Qin, Q. Zhang, C. Fang, J. Hu, Phys. Rev. B {\bf 98}, 104515 (2018).

\bibitem{Huang2019}
T. Huang, L. Zhang, and T. Ma, Sci. Bull. {\bf 64}, 310 (2019).



\bibitem{Wu2010}
C. Wu and J. Hirsch, Phys. Rev. B {\bf 81}, 020508 (2010).




\bibitem{Wang2014}
Y. Wang and A. Chubukov, Phys. Rev. B {\bf 90}, 035149 (2014).

\bibitem{Wang2017}
Y. Wang and L. Fu, Phys. Rev. Lett. {\bf 119}, 187003 (2017).

\bibitem{Yang2017}
W. Yang, C. Xu, and C. Wu, arXiv:1711.05241 (2017).

\bibitem{Lee2009}
W.-C. Lee, S.-C. Zhang, and C. Wu, Phys. Rev. Lett. {\bf 102}, 217002 (2009).

\bibitem{Hu2020}
L. H. Hu, P. D. Johnson, C. Wu
Phys. Rev. Research {\bf 2}, 022021(R)

\bibitem{Thomale2011}
R. Thomale, C. Platt, W. Hanke, and B. A. Bernevig, Phys. Rev. Lett. {\bf 106}, 187003 (2011).

\bibitem{Platt2012}
C. Platt, R. Thomale, C. Honerkamp, S.-C. Zhang, and W. Hanke, Phys. Rev. B {\bf 85}, 180502 (2012).

\bibitem{Khodas2012}
M. Khodas and A. V. Chubukov, Phys. Rev. Lett. {\bf 108}, 247003 (2012).

\bibitem{Fernades2013}
R. M. Fernandes and A. J. Millis, Phys. Rev. Lett. {\bf 111}, 127001 (2013).

\bibitem{Hinojosa2014}
A. Hinojosa, R. M. Fernandes, and A. V. Chubukov, Phys. Rev. Lett. {\bf 113}, 167001 (2014).

\bibitem{Lin2016}
S.-Z. Lin, S. Maiti, and A. Chubukov, Phys. Rev. B {\bf 94}, 064519 (2016).

\bibitem{Stanev2010}
V. Stanev and Z. Te\v sanov\'ic, Phys. Rev. B {\bf 81}, 134522 (2010).

\bibitem{Lin2012}
S.-Z. Lin and X. Hu, Phys. Rev. Lett. {\bf 108}, 177005 (2012).

\bibitem{Marciani2013}
M. Marciani, L. Fanfarillo, C. Castellani, and L. Benfatto, Phys. Rev. B {\bf 88}, 214508 (2013).

\bibitem{Maiti2013}
S. Maiti and A. V. Chubukov, Phys. Rev. B {\bf 87}, 144511 (2013).

\bibitem{Ahn2014}
F. Ahn, I. Eremin, J. Knolle, V. B. Zabolotnyy, S. V. Borisenko, B. B\"uchner, and A. V. Chubukov, Phys. Rev. B {\bf 89}, 144513 (2014).

\bibitem{Garaud2014}
J. Garaud and E. Babaev, Phys. Rev. Lett. {\bf 112}, 017003 (2014).

\bibitem{Maiti2015}
S. Maiti, M. Sigrist, and A. Chubukov, Phys. Rev. B {\bf 91}, 161102 (2015).


\bibitem{Garaud2011}
J. Garaud, J. Carlstr\"om, and E. Babaev, Phys. Rev. Lett. {\bf 107}, 197001 (2011).

\bibitem{Garaud2013}
J. Garaud, Johan Carlstr\"om, E. Babaev, and M. Speight, Phys. Rev. B {\bf 87}, 014507 (2013).

\bibitem{Lin2014}
S.-Z. Lin,  J. Phys.: Condens. Matter {\bf 26}, 493202 (2014).

\bibitem{Yerin2014}
Y. S. Yerin and A. N. Omelyanchouk, Low Temp. Phys. {\bf 40}, 943 (2014).

\bibitem{Yerin2017}
Y. Yerin, A. Omelyanchouk, S.-L. Drechsler, D. V. Efremov, and J. v. d. Brink, Phys. Rev. B {\bf 96}, 144513 (2017).


\end{thebibliography}

\end{document}